
\input phyzzx



\nopubblock
\PHYSREV
\parindent=0 truecm
\parskip=0 truecm
\titlepage
{\bf
\title{A STANDARD MODEL SOLUTION TO THE SOLAR NEUTRINO PROBLEM?\foot{
Supported in part by the
Technion Fund For Promotion of Research } }
\author{Arnon Dar and Giora Shaviv}
\address{Department of Physics and Asher Space Research Institute,
Technion-Israel Institute of Technology, Haifa 32000, Israel.}}
\abstract
An imroved standard solar model, with more accurate input parameters
and more accurate treatment of plasma physics effects,
predicts a solar neutrino flux which is consistent within experimental
uncertainties with the solar neutrino observations after 1986.

\endpage
The Sun is a typical main sequence star that
generates its energy by
the pp chain and CNO nuclear cycle\Ref\DC{
D. Clayton, ``Principles of Stellar Evolution And Nucleosynthesis'',
McGraw-Hill 1968.}, which also produce neutrinos. These neutrinos
have been observed by four pioneering solar neutrino
($\nu_\odot$) experiments,
the Chlorine experiment at Homestake, the Cerenkov light water
experiment at Kamiokande, the Soviet-American Gallium
experiment (SAGE) at Baksan and the European Gallium experiment
(GALLEX) at Grand Sasso, but the measured $\nu_\odot$ fluxes
are significantly  below those predicted by Standard Solar
Models\Ref\SSM{For recent reviews see
J.N. Bahcall and M. Pinsonneault, Rev. Mod. Phys. {\bf 64}, 885 (1992);
S. Turck-Chieze and I. Lopes, Ap. J. {\bf 408}, 347 (1993). } (SSM):
The Cl experiment observed over 20 years an average
production rate of\Ref\DL{R.
Davis Jr. et al. ``Frontiers of Neutrino Astrophysics'', eds. Y.
Suzuki and K. Nakamura (Universal Academic Press, Inc. Tokyo,
Japan, 1993) p. 47;
K. Lande, Bull. Am. Phys. Soc. {\bf 38}, 1797 (1993).}
$2.28\pm 0.23~SNU$ ($SNU=10^{-36}$ captures/atom s) of
$^{37}$Ar by solar $\nu_e$'s with energies above $0.81~MeV$
while the SSM predicts $7.4\pm 2.8~SNU$ (unless otherwise stated, we will
use the predictions of Bahcall and Pinsennault$^{^2}$ (BP) as a standard
reference).
Kamiokande has observed since 1987 electron recoils with
energy above $ 7.5~MeV$ from scattering of
$\nu_\odot$'s during 1040 (Kamiokande II) and 395 (Kamiokande III)
days of detector live time with event rates that are\Ref\KAMIO{K.
Nakamura, Nucl. Phys. B (Suppl.) {\bf 31}, 105 (1993).} $51\%\pm 6\%$
and $56\%\pm 6\%$, respectively, of that predicted by the SSM.
GALLEX has measured
a capture rate of solar $\nu_e$'s by $^{71}Ga$ of\Ref\GALLEX{P.
Anselmann, et al., Phys. Lett. B {\bf 314}, 445 (1993).}
$ 87\pm 16~SNU~ $ in 21 runs between May 1991 and February 1993,
compared with $128^{+19}_{-16}~SNU,$ predicted by the SSM.
SAGE began its $\nu_\odot$ observations in 1990.
Its initial results indicated a large discrepancy with
the SSM predictions\Ref\SAGE{A.I. Abazov et al., Phys. Rev. Lett.
{\bf 67}, 3332 (1991).}. However, the capture rate observed by SAGE
in 1991\Ref\SAGEII{V.N. Gavrin et al.,  Proc. XXVI ICHEP, ed. J Sanford,
(AIP, New York 1992) p. 1093},
$85^{+22}_{-32}[stat]\pm 20[sys]~SNU,$ agrees
with that observed by GALLEX. In spite of impressive
theoretical and experimental efforts the origin of these discrepancies,
which became known as the $\nu_\odot$ problem, is still unknown.

Bahcall and Bethe\Ref\BB{J.N. Bahcall and H. A. Bethe, Phys. Rev.
D {\bf 47}, 1298 (1993).} have argued that the solution of this
$\nu_{\odot}$ problem requires new physics
beyond the Standard Electroweak
Model because the Cl detector with an energy threshold of 0.81 MeV
observes a smaller solar $\nu_e$ flux than that
observed by Kamiokande whose energy threshold is 7.5 MeV.
However, after the instalment of new pumps in 1986,
the average capture rate measured by the Cl detector
is not in a serious conflict with the $\nu_\odot$ flux measured by
Kamiokande during the same period (see Fig.1), especially in
view of the absence of direct calibration of the Cl experiment.
Moreover, the $\nu_{\odot}$ spectrum measured by Kamiokande II
and III is consistent with the $\nu_e$ spectrum of $^8$B, as predicted
by the SSM. Furtheremore, the observed capture rates in GALLEX and
SAGE are those predicted directly from the solar luminosity and
the observed fluxes of the energetic $\nu_{\odot}$'s
in the Homestake and Kamiokande experiments,
using essentially only general conservation laws\Ref\BAKSAN{A. Dar,
Proc. Baksan 1993 Intl. School on Particle Physics and Cosmology,
Baksan, Russia (April 1993, in Press).}:
Since the net outcome  in the pp and CNO cycles is the
conversion of protons into Helium
nuclei, conservation of baryon number, charge
lepton flavour and energy requires that
   $ 4p+2e^-\rightarrow {^4He}+2\nu_e+Q~ ,$
where $Q\approx 26.73~MeV,$ i.e.,
two $\nu_e$'s are produced in the Sun
per 26.73 MeV release of nuclear energy. Thus,
if the Sun is approximately in a steady state where its nuclear
energy production rate equals its luminosity
 then the $\nu_{\odot}$ flux at Earth is given by
$$
\phi_{\nu_\odot}= {2L_{\odot}\over Q-2\bar{E}_{\nu}}~{1\over 4\pi D^{2}}
           \approx  6.46\times 10^{10}~cm^{-2} s^{-1}~, \eqno\eq $$
where $L_{\odot}=3.826(8)\times 10^{33}~erg~s^{-1}$ is the Sun's
luminosity, $D\approx 1.496 \times 10^{13}~cm$ is its distance from Earth
and $\bar {E}_\nu\approx \bar{E}_\nu(pp)\approx 0.265~MeV$
is the average $\nu_\odot$ energy (Kamiokande and the Cl experiments
observe much smaller fluxes of the more energetic solar neutrinos).
If the suppression of the flux of $\nu_\odot$'s other than the
pp neutrinos is energy independent and equals that observed
at Homestake, $S\equiv \phi_{\nu_\odot}/ \phi_{SSM}\approx 0. 31$, then
the predicted capture rate in  Gallium is
$<\phi\sigma>_{Ga}\approx 77-6S+57S ~SNU
\approx 93~SNU, $ where $77-6S~SNU$ is the capture rate of the pp
$\nu_\odot$'s and $57S~SNU$ is the
capture rate of all other $\nu_\odot$'s.
If the suppression is $S\approx 0.51$, as observed in Kamiokande,
then $<\phi\sigma>_{Ga}\approx 103~SNU$. Both predictions  are
in good agreement with the observations of GALLEX and SAGE.
All these suggest that perhaps the difference between
the results of the Cl experiment prior to 1986 and those
of Kamiokande after 1986 are due to statistical fluctuations
or systematic errors, allowing a standard physics solution to the
$\nu_\odot$ problem.

Indeed, in this letter we show that the SSM predictions
are consistent with the $\nu_\odot$ observations of Kamiokande,
SAGE and GALLEX. This is achieved mainly by:
(1) using an improved numerical stellar evolution
code\Ref\KOVETZ{A. Kovetz and G. Shaviv, Ap. J. May 1, 1994.}
to calculate the evolution of the Sun
from its pre-main sequence phase to the present stage, (2) using
an updated solar luminosity\Ref\PDP{Particle Data Group,
Phys. Rev. D {\bf 45}, III.2 (1992).}
$L_{\odot}=3.826(8)\times 10^{33}~erg~s^{-1}$
instead of $3.86\times 10^{33}~erg~s^{-1}$ used before,
which reduces the $^7$Be and $^8$B  $~\nu_\odot$ fluxes
by 3\% and 6\%, respectively, (3) using the updated values
$S_{34}(0)=0.45~keV\cdot b$ and $S_{17}(0)=17~ eV\cdot b$
for the low energy nuclear reactions $^3$He$(\alpha,\gamma)^7$Be
and $^7$Be(p,$\gamma)^8$B, respectively, instead of
$S_{34}(0)=0.56~keV\cdot b$ and $S_{17}(0)=22.4~eV\cdot b$,
(4) suppressing the overestimated
screening enhancement of bare nuclear cross sections in the
solar plasma (induces only a minor change after the above changes).

The rate of the neutrino producing reactions in the Sun
depends on the present conditions inside the Sun.
These are obtained from calculating the
evolution of the Sun from its pre-main sequence fully convective stage
to its present ($t_{\odot}\approx 4.55~GY$) stage. Its initial
composition, which is not known accurately enough, is adjusted
by requiring that the calculations yield the presently
observed properties of Sun, i.e.,
its radius, luminosity,
observed surface elemental abundances
and an internal structure consistent with helioseismology
data\Ref\CD{See e.g., J. Christensen-Dalsgaard and W.
Dappen, Astron. Astrophys. Rev. {\bf 4}, 267 (1992).}.
The above scheme with the best available input physics
(equations of state, opacities, nuclear cross sections)
has been improved and updated
continuously by Bahcall and his collaborators and more recently also
by other groups$^2$. Our predictions below were obtained
with an improved stellar evolution code$^{^{10}}$
which contains diffusion and partial ionization of {\bf all} elements,
follows the full evolution of the Sun
from its pre-main sequence stage until present and does not impose
nuclear equilibrium at any stage. The code uses the thermonuclear
reaction rates compiled by Caughlan and Fowler\Ref\CF{
G.R. Caughlan and W.A. Fowler, Atomic Data and Nuclear Data
Tables {\bf 40}, 284 (1988)} except for the changes explained
below.

{\bf Low energy nuclear cross sections} in the SSM were
extrapolated from measurements at much higher energies,
using the parametrization,
$$ \sigma(E)= {S(E)\over E}e^{-(E_G/E)^{1/2}}  \eqno\eq $$
where $E_G\equiv  2(\pi \alpha Z_1Z_2)^2\mu c^2, $
$Z_1$ and $Z_2$ being the charge numbers of the
colliding nuclei, $\mu$ their reduced mass
and $E$ their center-of-mass energy. $S(E)$ was assumed
to vary slowly with energy .
However, the above parametrization ignores the non negligible
contribution$^{^{1,}}$\Ref\RTC{R.G.H.
Robertson, Phys. Rev. C {\bf 7}, 543 (1973); S. Turck - Chieze
et al., Ap. J. {\bf 335}, 415 (1988).}
from partial waves other than s wave  at the high
energies where the cross sections have been measured
and the effect of a finite nuclear radius $R$
on barrier penetration, which induce an energy dependence in $S(E)$.
Indeed, $S(E)$ was found to be energy dependent for some key
reactions and its extrapolation to low energies became model dependent.
However, the use of the correct  Coulomb barrier penetration
factor for finite size nuclei and the inclusion of higher partial
waves remove most of the energy dependence of $S(E)$. In particular,
for effective potentials  $V=-V_0\leq0$ for $r<R$ and
$V=Z_1Z_2e^2/r$ for $r\geq R$,
the induced energy dependence of $S(E)$
well below the Coulomb barrier and far from resonance energies is
given by
$$ S(E)\approx \bar{S}e^{-
{2\over\pi}\left ({E_G\over E}\right)^{1/2}[2\sqrt{x}-arcsin\sqrt{x}
  -\sqrt{x(1-x)}]}~, \eqno\eq $$ where
$\bar{S}=S(0)+S'(0)E$, $x=E/E_c$ and $E_c=Z_1Z_2e^2/R~.$
That is demonstrated in Fig.2 for the reaction
$^3$He$(\alpha,\gamma)^7$Be, where we plotted $\bar{S}_{34}(E)$
as function of $E$, using $R=R(^3$He)+$R(\alpha)$=3.82 fm and
the experimental data of Parker and
Kavanagh\Ref\PK{P.D. Parker and R.W. Kavanagh, Phys. Rev. {\bf 131},
2578 (1963).} and of Osborne et al.\Ref\OET{J.L. Osborne et al., Nucl.
Phys. A {\bf 419}, 115 (1984).} Indeed,
$\bar{S}_{34}$ shows very small energy dependence at low enough energies
where the contribution from higher partial waves is negligible.
With this form for the barrier penetration factor, the world experimental
data on the reaction $^3$He$(\alpha,\gamma)^7$Be yield
$S_{34}(0)=0.45~\pm 0.03~keV\cdot b$, which is significantly smaller than
the value $S_{34}(0)=0.56~keV\cdot b$ used by Bahcall and
Pinsennault$^{^2}$.

Using the experimental data of Filippone et al.\Ref\FET{B.W. Filippone
et al., Phys. Rev. Lett. {\b 50}, 412 (1983); Phys. Rev. C {\bf 28},
2222 (1983); B.W. Filippone, Ann. Rev. Nucl. Sc. {\bf 36}, 717 (1986).}
on the reaction $^7$Be(p,$\gamma)^8$B at low energies, where the
contribution from higher partial waves is small,
and the recent measurements by Motobayashi et al.\Ref\MET{
T. Motobayashi et al., Phys. Rev. Lett., Submitted.}
of $S_{17}(E)$ at higher energies from Coulomb dissociation of $^8$B,
which are free of the resonating p-wave and
higher partial waves that contribute
to the law energy cross section of
$^7$Be(p,$\gamma)^8$B, we find that for
$2.4~fm\leq R(p)+R(^7$Be)$\leq 3.2~fm$,
$\bar{S}_{17}$ is practically energy independent
and  $S_{17}(0)=17\pm 2 ~ eV\cdot b$ (see Fig. 3).
This value which we adopt was
advocated before by Barker and Spear\Ref\BS{F.C. Barker, and R.H. Spear,
Ap. J. {\bf 307}, 847 (1986).} and it is significantly smaller than
$S_{17}(0)=22.4\pm 2 ~ eV\cdot b$ used by Bahcall and Pinsennault$^{^2}$.

{\bf Screening} of the Coulomb potential of
target nuclei by their electrons is known to enhance
significantly laboratory nuclear cross sections at very low
energies\Ref\ENG{S. Engstler et al., Phys. Lett. B {\bf 202}, 179
(1988).}, although a complete theoretical understanding of the effect
is still lacking\Ref\SH{
T.D. Shoppa et al., Phys. Rev. C {\bf 48}, 837 (1993).}.
Screening corrections to the nuclear reaction rates in the SSM are
usually represented by an enhancement factor
$$ F\approx e^{\Delta U/kT}\approx e^{Z_iZ_je^2/R_DkT}~,\eqno\eq $$
where $\Delta U$ is the gain in electronic
potential energy when an incident ion of charge $Z_je$ penetrates
the electronic cloud of an ion of charge $Z_ie$, T is the plasma
temperature and $R_D$ is the Debye length,
$$R_D\equiv \left ({kT\over 4\pi e^2\Sigma Z^2 \bar n_{Z}}\right )~.
\eqno\eq $$
The change in the Coulomb barrier due to screening
is estimated from the Debye-Huckel approximation to the
screened potential around a static ion in an electrically neutral plasma,
($\Sigma Z\bar n_{Z}=0$,  $\bar n_{Z}$ being
the number density of particles of charge Ze with Z=-1 for electrons):
$$\Phi_i={eZ_i\over R_D}e^{-r/R_D}\approx {eZ_i\over r}-
   {eZ_i\over R_D} ~~{\rm for}~r\ll R_D. \eqno\eq $$
This potential is an approximate
solution to Poisson's equation near an ion of charge $Z_i$,
$$ \nabla^2\Phi_i=4\pi e\Sigma Z n_{Z}=
4\pi e\Sigma Z\bar n_{Z}e^{-eZ\Phi_i/kT}\approx R_D^2\Phi_i,\eqno\eq $$
which is valid, however,
only far from the nucleus where $eZ\Phi_i\ll kT~.$
In the core of the Sun where $kT\sim 1~keV$
is much smaller than the Coulomb barriers ($\sim 1~ MeV$) between
the reacting nuclei, most of the contribution to the nuclear
reaction rates, comes from collisions with c.m. energies
$E\gg kT~.$ At the classical turning point $eZ_j\Phi_i=E~.$
Consequently, inside the barrier, $eZ_j\Phi_i\gg kT$,
and the use of the Debye-Huckel approximate potential
at distances shorter than the classical turning point for
calculating the barrier penetration factor$^{^1}$ is unjustified.
The Debye-Huckel solution also requires that the inter-ion spacing
is much shorter than the Debye length, and that there are many
electrons within a Debye sphere. Both conditions are not satisfied
in the core of the Sun. It was found that when the conditions for the
validity of the Debye-Huckel approximation are badly violated in
laboratory plasmas, the Debye-Huckel screened
potential fails dramatically\Ref\GGC{S. Goldsmith et al., Phys. Rev.
A {\bf 30}, 2775 (1984).} in reproducing various plasma properties.

Moreover, an ion that approaches from infinity a nucleus
screened by an electronic cloud does gain electronic
potential energy $\Delta U\approx e^2Z_iZ_j/ R_D$
when it penetrates this cloud. But, there is no gain in potential
energy if either its
initial position is already inside the electronic cloud
(because the potential  $\sim eZ_i/ R_D~,$ is
constant there), or the ion leaves and enters similar potential wells.
Near the center of the Sun, where
$\rho_c\approx 156~g~cm^{-3},$ $T_c \approx 1.57\times 10^7K$,
$X_c(H)\approx 34\%$, and $X_c(He)\approx 64\%$,
the average interion spacing is $n_i^{-1/3}\sim 2.8\times 10^{-9}~cm,$
similar to the Debye length, $R_D\approx 2.3\times 10^{-9}~cm.$
Furthermore, the effective energies of the reacting ions are much higher
than kT and their velocities are similar to the thermal electron
velocities, not allowing sufficient time for the plasma to readjust
itself and screen effectively these ions. Consequently, both effects
reduce substantially
the energy gain due to screening and the resulting enhancement of
the fusion reaction rates near the center of the Sun.
The overestimated enhancement of {\bf all}
nuclear cross sections by screening
in the SSM has a small net effect on the Sun
but may be important for other stars and will be further investigated
in numerical simulations of stellar like plasmas\Ref\ADGS{A.
Dar and G. Shaviv, to be published.}.

Our SSM predictions for the solar neutrino fluxes, obtained
from the stellar evolution code of Kovetz and Shaviv$^{^{10}}$
with the above modifications
are summarized in Table I. They are consistent, within experimental
uncertainties, with the experimental results of Kamiokande, GALLEX
and SAGE, and of Homestake after 1986.
In particular, we predict
$\phi_\nu(^8$B)=2.77$\times 10^6~cm^{-2}s^{-1}$ in excellent
agreement with the flux measured by Kamiokande II and III,
but we do not explain why
the $\nu_\odot$ flux measured by the Cl experiment
prior to 1986 is smaller than the $\nu_\odot$ flux
measured later by Kamiokande.
A full account of our calculations will be published elsewhere.

{\bf Table I:} Comparison between experimental results$^{^{3-7}}$
and the predictions of the SSM of Bahcall and Pinsennault$^{^2}$ (BP),
Turck-Chieze and Lopes$^{^2}$ (LP), Kovetz and Shaviv$^{^{10}}$ (KS)
and our predictions (DS) as described in the text, for
the solar neutrino fluxes ($cm^{-2}s^{-1}$) at Earth
and their capture rates (SNU) in $^{37}$Cl and $^{71}$Ga.
$$\matrix{{\rm \nu_\odot~Flux}& BP   & TL & KS & DS & EXP         \cr
{\rm pp~(E10)}   & 5.95 & 6.02  & 5.99  & 6.04 &             \cr
{\rm pep~(E8)}   & 1.41  & 1.39 & 1.38  & 1.40 &             \cr
{\rm ^7Be~(E7)}  & 4.75  & 4.34 & 4.91  & 4.30 &             \cr
{\rm ^8B~~(E6)}   & 5.24  & 4.63 & 5.83  & 2.77 & 2.87\pm 0.17\cr
{\rm ^{13}N~(E7)}& 43.9  & 38.3 & 6.11  & 7.47 &             \cr
{\rm ^{15}O~(E7)}& 37.5  & 31.8 & 1.78  & 2.17 &             \cr
{\rm ^{17}F~(E6)}& 4.72  &      & 4.85  & 5.21 &             \cr
<\phi\sigma>_{Cl}& 7.4 & 6.4& 7.6&  4.2        & 2.28\pm 0.23\cr
<\phi\sigma>_{Ga}& 128 & 123 & 123 & 109       & 87\pm 16    \cr} $$
\endpage
\refout

\centerline{{\bf FIGURE CAPTIONS}}

{\bf Fig.1}:  Comparison between the observed rates
induced by solar neutrinos in the Cl (open circles)
and the Kamiokande II (square boxes)
experiments divided by the rates predicted
by the SSM of Bahcall and Ulrich (BU) as function of time
during the period January 1987 through April 1990. The dotted line
is the average of the Cl data points, the full line is the average of the
Kamiokande II data points.

{\bf Fig.2}: $\bar{S}_{34}$ as function of $E_{cm}$ as extracted
from the cross section for $^3$He$(\alpha,\gamma)^7$Be
measured by Parker and Kavanagh$^{^{15}}$ (triangles) and by
Osborne et al.$^{^{16}}$ (full circles). The straight line
represents the average $\bar{S}_{34}$ obtained from the world data
(not shown here).

{\bf Fig.3}: $\bar{S}_{17}$ as function of $E_{cm}$ as extracted
from the low energy cross section for $^7$Be(p,$\gamma)^8$B
measured by Filippone et al.$^{^{17}}$ (triangles) and from Coulomb
dissociation of $^8$B measured recently by Motobayashi et al.$^{^{17}}$
(full circles).
\endpage
\end